\documentclass[12pt,aps,prc,showpacs,amsmath,showkeys]{revtex4}

\usepackage[dvips]{graphicx}
\usepackage{epstopdf}
\usepackage{amssymb}
\usepackage{epstopdf}

\newcommand{\ba}{\begin{eqnarray}}
\newcommand{\ea}{\end{eqnarray}}
\newcommand{\bmath}{\begin{mathletters}}
\newcommand{\emath}{\end{mathletters}}
\newcommand{\ban}{\begin{eqnarray*}}
\newcommand{\ean}{\end{eqnarray*}}

\draft

\begin{document}

\title{Pseudospin Conserving Shell Model Interactions}
\author{Joseph N. Ginocchio}
\address{MS B283, Theoretical Division, Los Alamos National Laboratory, Los
Alamos, New Mexico 87545, USA}
\date{\today}

\begin{abstract}
Pseudospin symmetry is approximately conserved in nuclei. Normally shell model 
interactions are written in terms of spin an orbital angular momentum 
operators, not in 
terms of pseudospin and pseudo-orbital angular momentum operators. We determine 
the shell model interactions which conserve pseudospin symmetry and 
pseudo-orbital 
angular momentum symmetry and write them in terms of spin and orbital angular 
momentum operators including the tensor interaction. We show that, although the 
tensor interaction by itself does not conserve pseudo-orbital angular momentum, 
certain combinations of the tensor interaction with the two body angular momentum squared interaction and the two body spin orbit interaction do conserve pseudo-orbital angular momentum.
\end {abstract}



\maketitle

\newpage
\section{Introduction}

Pseudospin symmetry is a relativistic symmetry of the Dirac Hamiltonian that occurs when the scalar potential plus a constant is equal in magnitude to the vector potential but opposite in sign \cite {gino97}. This condition approximately holds for the relativistic mean fields of nuclei \cite {madland}. Indeed, nuclear energy levels and transition rates in both spherical and deformed nuclei are consistent with approximate pseudospin symmetry \cite {gino05}. Beyond the mean field the non-relativistic shell model with effective interactions has been very successful in describing nuclei. However, 
the  bare nuclear interaction and the effective shell model interactions between nucleons are expressed in terms of spin operators and not pseudospin operators. In this paper we shall determine the interactions which conserve pseudospin symmetry. These interactions are written in momentum space rather than coordinate space because pseudospin involves the intertwining of spin and momentum. However, this may be an advantage since recent effective interactions, including effective field theories with and without pions \cite {eft} and the low-k effective interactions \cite {low}, are written terms of momentum.

\section{Pseudospin Symmetry}

Pseudospin symmetry is an SU(2) symmetry as is spin symmetry. The relativistic generators for the pseudospin algebra,
${\tilde{S}}_{i,k} (i=x,y,z)$, where $k$ is the nucleon number, are \cite{ami}
\begin{equation} 
{\tilde S}_{i,k} = \left(\begin{array}{cc} 
{\tilde s}_{i,k} & \quad 0 \\[2pt]
0 & \quad s_{i,k}\end{array}\right)= \left(\begin{array}{cc}
U_p\, {s}_{i,k}U_p & \quad 0 \\[2pt]
0 & \quad {s}_{i,k}
\end{array}\right)\,,
\label{psg}
\end{equation}
where ${ s}_{i,k}=\sigma_{i,k}/2$ are the usual spin generators, $\sigma_i$
the Pauli matrices, and $U_p={\sigma_k}\cdot {\hat p}$ is the
momentum-helicity unitary operator \cite{draayer},  and ${\hat p}_{i,k}= {{ p_{i,k}}\over p}$ is the unit three momentum of a single nucleon. The four by four nature of the generators results from the fact that they are  relativistic generators. The generators for the non-relativistic pseudospin algebra are
\begin{equation}
{\tilde s}_{i,k} = U_p\ {  s}_{i,k}\ U_p = 2\ s_k \cdot  {\hat p}\ {\hat p}_{i,k} - s_{i,k}.
\label{gen1}
\end{equation}
We note that, although the pseudospin generators depend on momentum, they depend on the unit vector of momentum and therefore are equivalent to spin as far as momentum power counting in effective field theory.

%



For the two nucleons, the total non-relativistic pseudospin is 
\begin{equation}
{\tilde s}_{i} = {\tilde s}_{i,1}+ {\tilde s}_{i,2}= U_{1,p_1}\ { s}_{i,1}\ U_{1,p_1}+U_{2,p_2}\ { s}_{i,2}\ U_{2,p_2}= U_{1,p_1}U_{2,p_2}\ { s}_{i} \ U_{1,p_1}U_{2,p_2}
\label{gen}
\end{equation}
where the total spin is ${ s}_{i}= {s}_{i,1}+ {s}_{i,2}$.
In the rest frame $p_{i,1}= p_{i}, p_{i,2} = -p_{i}$, where $ {p_{i}}$ is the relative momentum. Hence the generators are
\begin{equation}
{\tilde s}_{i} = U_{1,p}U_{2,p}\ { s}_{i}\ U_{1,p}U_{2,p} =  2\ s \cdot  {\hat {p}}\ {\hat {p}}_{i }- s_{i}
\label{gent}
\end{equation}

Likewise the non-relativistic pseudo-orbital angular momentum is \cite{gino05}
\begin{equation}
{\tilde \ell}_{i} = {\sigma_1}\cdot  {\hat p}\   {\sigma_2}\cdot  {\hat p}\ {{\ell}_i}\  {\sigma_1}\cdot  {\hat p}\   {\sigma_2}\cdot  {\hat p}
\label{oam}
\end{equation}
where the orbital angular momentum is ${\ell}_i ={ (r \times p)_i \over \hbar}$ where $r$ is the relative coordinate.

\section{Pseudospin Symmetry Conserving Interaction}

The general interaction which conserves pseudospin, pseudo-orbital angular momentum, and the total angular momentum, $j_i = {\tilde {\ell}}_i + {\tilde s}_i$, is

\begin{eqnarray}
&~
 V(p) = \nonumber \\
 &({\tilde V}_{c}^{(0)}(p) +{\tilde V}_{ps}^{(0)}(p){\tilde s}_{1}\cdot {\tilde s}_{2} +{\tilde V}_{po}^{(0)}(p) {\tilde {\ell}}\cdot{\tilde {\ell}} +{\tilde V}_{pso}^{(0)}(p)  {\tilde s}\cdot {\tilde {\ell}}){(1-\tau_1\cdot \tau_2)\over 4}
\nonumber\\
&
+( {\tilde V}_{c}^{(1)}(p) +{\tilde V}_{ps}^{(1)}(p){\tilde s}_{1}\cdot {\tilde s}_{2} +{\tilde V}_{po}^{(1)}(p) {\tilde {\ell}}\cdot{\tilde {\ell}} +{\tilde V}_{pso}^{(1)}(p)  {\tilde s}\cdot {\tilde {\ell}}){(3+\tau_1\cdot \tau_2)\over 4},\nonumber\\
&
\label{psi}
\end{eqnarray}
where $ \tau$ are isospin Pauli matrices and we include the possibility that the coefficients $V_{A}^{(T)}(p)$, $A = c, ps, po, pso$, could depend on isospin, $T=0,1$. 

If $V_{pso}^{(T)}(p)$ = 0 then pseudospin and pseudo-orbital angular momentum are invariant symmetries; that is, the eigenfunctions have conserved pseudospin and pseudo-orbital angular momentum quantum numbers and the energies do not depend on the orientation of the pseudospin. In this case the generators in  Eq (\ref {gent}) and Eq (\ref {oam}) commute with the interaction. If $V_{pso}^{(T)}(p) \ne 0$ then pseudospin and pseudo-orbital angular momentum are dynamical symmetries; that is, the eigenfunctions have conserved pseudospin and pseudo-orbital angular momentum quantum numbers but the energies are not degenerate. In this case the generators in   Eq (\ref {gent}) and Eq (\ref {oam}) do not commute with the interaction. This is the most realistic possibility.

\section{Pseudo-Operators in terms of Normal Operators}

Since nuclear interactions are usually written in terms of spin and orbital angular momentum (normal operators), we rewrite the pseudospin and pseudo-orbital angular momentum operators (pseudo-operators) in the interaction in Eq (\ref {psi}) in terms of these normal operators.

First we consider the pseudospin-pseudospin interaction
\begin{equation}
{\tilde s}_1\cdot {\tilde s}_2 = ( \sigma_1 \cdot  {\hat {p}}\ {\hat {p}}- s_{1})\cdot ( \sigma_2 \cdot  {\hat {p}}\ {\hat {p}}- s_{2})=  \sigma_{1} \cdot  {\hat {p}}\ \sigma_{2 }\cdot  {\hat {p}}-
\sigma_{2} \cdot  {\hat {p}}\ s_{1}\cdot  {\hat {p}} -  \sigma_1 \cdot  {\hat {p}}\  s_{2}\cdot  {\hat {p}} +  { s}_{1}\cdot { s}_{2},\label{int}
\end{equation}

which leads to
\begin{equation}
{\tilde s}_1\cdot {\tilde s}_2 = { s}_1\cdot { s}_2;
\label{pspin}
\end{equation}
that is, the two are equivalent. This is consistent with the study of the nucleon-nucleon interaction \cite {gino02} in which it was shown that the pseudospin transformation on two nucleons does not change the spin. However, the mixing angle between states with the same pseudospin but different pseudo-orbtial angular momentum can be different than the mixing angle between states of with the same spin but different orbital angular momentum, which comes about through other terms involving the pseudo-orbital angular momentum operator and the orbital angular momentum operator.

Furthermore, from Eq.(\ref{gen}), the tensor interaction becomes
\begin{equation}
\sigma_1\cdot {p}\   \sigma_2\cdot  {p}\ =p^2\  [{\tilde s}_1\cdot { s}_2 + {s}_1\cdot {\tilde s}_2 +{\tilde s}_1\cdot {\tilde s}_2+{ s}_1\cdot {s}_2]. 
\label{t}
\end{equation}

That is, the tensor interaction is symmetrical in pseudospin and spin and it is an  interaction between the spin and pseudospin, which is an interesting insight.


We consider now interactions involving the pseudo-orbital angular momentum. 
The two body pseudospin-pseudo-orbit interaction becomes

\begin{equation}
{\tilde s} \cdot {\tilde \ell}= { -s \cdot  \ell}+{ \sigma_{1}\cdot  {\hat p} \ \sigma_{2}\cdot  {\hat p}} + 1 - 2\  s \cdot  s ,
 \label{T}
\end{equation}

This means that this interaction can be written in terms of the two body spin-orbit interaction and the tensor interaction. On the other hand the pseudo-orbital angular momentum squared is 

\begin{equation}
{\tilde  \ell}\cdot{\tilde  \ell} =  \ell\cdot  \ell + 4\  s\cdot  \ell - 2 +  4\  s\cdot  s - 2\
{\sigma_1}\cdot  {\hat p}\   {\sigma_2}\cdot  {\hat p}\
\label{ell2}
\end{equation}

which means that the pseudo-orbital angular momentum squared can also be written in terms of the the orbital angular momentum squared, two body spin-orbit interaction and the tensor interaction. So the pseudospin and pseudo-orbital angular momentum do not introduce any new terms that are not already present with the spin and orbital angular momentum. Hence the interaction in Eq.(\ref{psi}) can be written in terms of operators involving spin and orbital angular momentum.
\section{The Interaction in terms of Normal Operators}
Using the relations between peudo-operators and normal operators given in Eq.(\ref{pspin})-Eq.(\ref{ell2}) we can rewrite Eq.(\ref{psi}) in terms of normal operators. This interaction becomes
\begin{eqnarray}
&~
 V(p) = \nonumber \\
 &({ V}_{c}^{(0)}(p) +{ V}_{s}^{(0)}(p){ s}_{1}\cdot { s}_{2} +{ V}_{o}^{(0)}(p) { {\ell}}\cdot{  {\ell}} +{  V}_{so}^{(0)}(p)  {  s}\cdot {  {\ell}}+{ V}_{t}^{(0)}(p){\sigma_1}\cdot  {\hat p}\   {\sigma_2}\cdot  {\hat p}){(1-\tau_1\cdot \tau_2)\over 4}
\nonumber\\
&
+( {  V}_{c}^{(1)}(p) +{  V}_{s}^{(1)}(p){  s}_{1}\cdot {  s}_{2} +{  V}_{o}^{(1)}(p) {  {\ell}}\cdot{  {\ell}} +{  V}_{so}^{(1)}(p)  {  s}\cdot {  {\ell}} +{ V}_{t}^{(0)}(p){\sigma_1}\cdot  {\hat p}\   {\sigma_2}\cdot  {\hat p}){(3+\tau_1\cdot \tau_2)\over 4},\nonumber\\
&
\label{si}
\end{eqnarray}
with
\begin{equation}
{ V}_{c}^{(T)}(p) =  { \tilde V}_{c}^{(T)}(p) + {  \tilde V}_{pso}^{(T)}(p) - 2{ \tilde V}_{po}^{(T)}(p) \\
\end{equation}
\begin{equation}
{ V}_{s}^{(T)}(p) =  { \tilde V}_{ps}^{(T)}(p) + 4{ \tilde V}_{po}^{(T)}(p)  - 2{  \tilde V}_{pso}^{(T)}(p) \\
\end{equation}
\begin{equation}
{ V}_{o}^{(T)}(p) =   { \tilde V}_{po}^{(T)}(p) \\
\end{equation}
\begin{equation}
{ V}_{so}^{(T)}(p) =  4{ \tilde V}_{po}^{(T)}(p)-{  \tilde V}_{pso}^{(T)}(p) \\
\end{equation}
\begin{equation}
{ V}_{t}^{(T)}(p) =  { \tilde V}_{pso}^{(T)}(p)- 2{  \tilde V}_{po}^{(T)}(p) \\
\label{s}
\end{equation}
In particular the tensor interaction is part of the interaction which conserves pseudo-orbital angular momentum, although by itself the tensor interaction breaks pseudo-orbital angular momentum. The conservation comes with a certain combination of tensor interaction, two-body angular momentum squared, and two-body spin-orbit. In fact from the relations above we derive, 

\begin{equation}
{ V}_{t}^{(T)}(p) + {V}_{so}^{(T)}(p) - 2{V}_{o}^{(T)}(p) =0.
\label{t}
\end{equation}

This also implies that for
\begin{equation}
 {V}_{so}^{(T)}(p) = 2{V}_{o}^{(T)}(p), 
\label{t0}
\end{equation}
the tensor interaction vanishes and both spin and pseudospin are conserved.
\section{Summary}
We have shown that there exist shell model interactions with tensor interactions which conserve pseudospin and pseudo-orbital angular momentum. These interactions have a relation between the strength of the tensor interaction, the strength of the two-body orbital angular momentum squared, and the strength of two-body spin orbit which is given in Eq.(\ref{t}). These interactions include those for which pseudospin and pseudo-orbital angular momentum are dynamical symmetries, which are the most realistic interactions. 

The tensor interaction has been shown to be important for shell evolution in exotic nuclei \cite  {otsuka}. At the same time pseudospin doublets are also seen in these nuclei \cite {sorlin}. Perhaps the interactions discussed in this paper will be able to explain both effects in a unified way. 

This research was supported by the United States Department of
Energy under contract W-7405-ENG-36.\\

\end{document}